\documentclass[twocolumn,floatfix,amsfonts,amssymb,notitlepage]{revtex4-1}
\usepackage{color}
\usepackage{graphicx}
\usepackage{amsmath}
\usepackage{latexsym}
\usepackage{epstopdf}
\usepackage{amsmath}
\usepackage{amssymb}
\usepackage{dsfont}
\usepackage{multirow}
\usepackage{mathtools}
\usepackage{tcolorbox}
\usepackage{ulem}

\newcommand{\beq}{\begin{equation}}
\newcommand{\eeq}{\end{equation}}
\newcommand{\bea}{\begin{eqnarray}}
\newcommand{\eea}{\end{eqnarray}}

\newcommand{\bebox}{\begin{tcolorbox}}
\newcommand{\eebox}{\end{tcolorbox}}

\newcommand{\eq}{\begin{equation}}
\newcommand{\en}{\end{equation}}
\newcommand{\ear}{\begin{eqnarray}}
\newcommand{\rae}{\end{eqnarray}}

\newcommand{\rf}[1]{(\ref{#1})}
\newcommand {\inpar}[1]{\lfloor{#1}\rfloor}
\newcommand {\mbar}{\overline{M}}
\newcommand{\scp}{\scriptsize}

\newcommand{\id}{\mathds{1}}
\newcommand{\be}{\begin{eqnarray}}
\newcommand{\ee}{\end{eqnarray}}
\newcommand{\non}{\nonumber}

\makeindex

\begin{document}
\title{Free-parafermionic $Z(N)$  and free-fermionic $XY$ quantum chains}
\author{Francisco C. Alcaraz}
\email{alcaraz@ifsc.usp.br}
\affiliation{ Instituto de F\'{\i}sica de S\~{a}o Carlos, Universidade de S\~{a}o Paulo,
Caixa Postal 369, 13560-970, S\~{a}o Carlos, SP, Brazil}
\author{Rodrigo A. Pimenta}
\email{pimenta@ifsc.usp.br}
\affiliation{Departamento de F\'isica, 
          Universidade Federal de Lavras, 
          Caixa Postal 3037, 37200-000, Lavras, MG, Brazil}
\date{\today{}}

\begin{abstract}
 
The relationship between the eigenspectrum of Ising and 
XY quantum chains is well known. Although the Ising model has a $Z(2)$ symmetry and the XY 
model a $U(1)$ symmetry, both models are described in terms of free-fermionic 
quasi-particles. The fermionic quasi-energies are obtained by means of a 
Jordan-Wigner 
transformation. On the other hand, there exist in the literature a huge 
family of $Z(N)$ quantum chains whose eigenspectra, for $N>2$, are given 
in terms of free parafermions and they  are not derived  from the standard 
 Jordan-Wigner transformation. The first members of this family are the 
$Z(N)$ free-parafermionic Baxter quantum chains. In this paper we introduce 
a family of XY models that beyond two-body also have 
$N$-multispin interactions. Similarly to the standard  XY model they have a $U(1)$  
symmetry 
and are also solved by the Jordan-Wigner transformation. 
We show that with 
appropriate choices of the $N$-multispin couplings, the eigenspectra of 
these XY models are given in terms of combinations of $Z(N)$ free-parafermionic quasi-energies. 
In particular all the eigenenergies of the $Z(N)$ free-parafermionic models
are also present in the related free-fermionic XY models.
The correspondence is established via the identification of the
characteristic polynomial which fixes the eigenspectrum. In the $Z(N)$ free-parafermionic models the quasi-energies 
obey an exclusion  circle principle that is not present in the related 
$N$-multispin XY models.

\end{abstract}

\maketitle

\section{Introduction}

The first step toward understanding an interacting system is to 
consider the simplest models where the eigenenergies are given in terms of 
free-particle quasi-energies. These are the so-called free systems and they
play an important role in condensed matter, statistical physics and 
quantum information theory. 

The most studied free quantum chains are the quantum Ising model in a 
transverse field and the XY quantum chains, also known as XX quantum chains
 \cite{lieb,pfeuty}. They describe the 
dynamics of spin-1/2 quantum spins attached to the sites of a lattice. 
Although the Ising quantum chain has a simpler $Z(2)$ symmetry and the 
XY model has a larger $U(1)$ symmetry, the eigenspectra of both models are 
related. The $2^L$ eigenenergies of the  XY model with $L$ sites ($L$ 
even) are obtained from those of two decoupled Ising quantum chains 
with $L/2$ sites. 
 In \cite{abb} this correspondence is shown for the general case of
the Ashkin-Teller model and the XXZ quantum chain. These models for the
vanishing anisotropy give us the XY model and two decoupled Ising models.

An interesting extension of these free-fermionic models are the $Z(N)$
free-parafermionic Baxter chains
\cite{baxter1,baxter2,fendley1,Baxter2014,Au_Yang_2014,auyang2016parafermions,AB1,AB2}.
They are generalizations of the Ising models described in terms of $N\times N$ 
matrices satisfying a $Z(N)$ algebra. On a lattice of size $L$ the $N^L$ 
eigenenergies $E_{\scp{s_1,\ldots,s_L}}$ are obtained by combining $L$ 
distinct pseudo-energies $\varepsilon_k$ ($k=1,\ldots,L$):
\be \label{1.1}
E_{\scp{s_1,\ldots,s_L}} = -\sum_{i=1}^L e^{i2\pi s_k/N} \varepsilon_k\,, \; 
s_k=0,1,\ldots,N-1.
\ee

We have in \rf{1.1} a $Z(N)$-circle exclusion since any quasi-energy 
$\varepsilon_k$ enters with {\it one and only one } possible value $s_k=0,1,2,\ldots,N-1$.

The $N=2$ case recovers the free-fermionic quantum Ising chain, and the fermion exclusion 
principle translates into the $Z(2)$-circle exclusion. On each eigenlevel the 
appearance of the energy $\varepsilon_k$ excludes the appearance of 
$e^{i2\pi /2}\varepsilon_k = -\varepsilon_k$. In the related free-fermionic 
XY model, with $2L$ sites, the same quasi-energies $\varepsilon_k$ 
($k=1,\ldots,L$) appear, but now the energies may or may appear, independently,
and we do not have the $Z(2)$-circle exclusion observed in the quantum 
Ising chain. 

The aim of this paper is to introduce a new family of $U(1)$-symmetric 
XY models that extend the observed 
Ising-XY correspondence to the $Z(N)$ free-parafermionic Baxter chains, with 
$N>2$. Toward this end, we introduce an extension of the  XY model that 
beyond the usual two-body interactions also contains $N$-multispin interactions 
($N=2,3,\ldots$). The $N=2$ case recovers the standard XY model 
(or the XX model), but for 
$N>2$ the models have a free-fermionic eigenspectra with complex quasi-energies 
which are associated with the ones appearing in a $Z(N)$ free-parafermionic Baxter chain, but with
no $Z(N)$-circle exclusion. 
Although  they share eigenvalues with a quantum chain with $Z(N)$ 
symmetry, these models have an $U(1)$ symmetry, since the $z$-component of the total 
magnetization is a good quantum number.

Recently, a new family of $Z(N)$ free-parafermionic quantum chains with 
$(p+1)$-multispin interactions were introduced ($p=1,2,\ldots$) \cite{AP1,AP2}. 
In the $N=2$ case these multispin models are also related to a free-fermionic 
model in a frustration graph \cite{network}.
The $p=1$ case recovers the $Z(N)$ free-parafermionic Baxter chains. These 
models for $N>2$ are non-Hermitian and exhibit multicritical points 
belonging to new universality classes of critical behavior. 
In the cases where $N\geq (p+1)$ we were able to derive $N$-multispin XY quantum chains whose eigenspectra are given in terms of the quasi-energies of 
these new $Z(N)$ multispin models. 
That is, the quasi-energies forming the eigenspectra of both models are the
same. In particular, all the eigenlevels of these free parafermionic
multispin models are also present in the eigenspectra of the free-fermionic
XY quantum chains with N-multispin interactions.

The correspondences reported in this paper allow us to give the exact 
ground-state energy and the critical exponents at the multicritical points 
of the generalized multispin XY quantum chains. 

This paper is organized as follows. In the next section we review the 
$(p+1)$-multispin $Z(N)$ quantum chains introduced in \cite{AP1,AP2}. In 
Sec. III, we present the generalized XY models with $N$-multispin interactions. 
 In Sec. IV, we give the correspondence of the generalized XY models with the 
$Z(N)$ free-parafermionic Baxter chains. Finally, in Sec. V we present our conclusions. In  Appendix A, we extend  
the results of Sec. IV by giving the XY models related to the new 
families \cite{AP1,AP2} of $Z(N)$ free-parafermionic multispin models of Sec. II. In Appendix B, we provide some simple numerical comparisons of the 
quasi-energies of the multispin XY and $Z(N)$ quantum chains.

\section{General free-fermionic and free-parafermionic quantum chains}

Recently, in \cite{AP1,AP2}, a large family of quantum chains were introduced
whose eigenspectra are given in terms of  free quasi-particles. Here we collect some key results. On their 
general formulation these Hamiltonians are written as a  sum of $M$ generators 
$\{h_i\}$,
\be \label{2.1}
H_M^{(N,p)} (\lambda_1,\ldots,\lambda_M) = -\sum_{i=1}^M h_i, 
\ee 
where 
$N=2,3,\ldots$,  and $p=1,2,\ldots$. The generators satisfy
 the $Z(N)$ exchange algebra:
\bea \label{2.2}
&&h_ih_{i+m} = \omega h_{i+m}h_i \quad\mbox{ for } \quad 1\leq	m \leq p;\quad  \omega = e^{i2\pi/N}, 
\nonumber \\
 && [h_i,h_j]=0 \mbox{ for } |j-i|>p,
\eea
with the closure relation 
\be \label{2.3} 
h_i^N = \lambda_i^N, 
\ee
given in terms of the scalars $\{\lambda_i\}$. For any representation 
\rf{2.2} and \rf{2.3} with integer values of $N\geq2$ and $p\geq1$ the Hamiltonian has a 
free-particle spectrum. For $N=2$ we have a free-fermionic spectrum and for 
$N>2$ we have a free-parafermionic one.

The standard quantum Ising model in a lattice with $L=(M+1)/2$ sites, and 
arbitrary couplings $\{\lambda_i\}$ give us a representation for $N=2$, $p=1$ 
and odd $M$, namely
\bea \label{2.4}
&&H_{2L-1}^{(2,1)}(\lambda_1,\ldots,\lambda_{2L-1}) = -\sum_{i=1}^L 
 \lambda_{2i-1}\sigma_i^x \non \\&&\quad\quad\quad- \sum_{i=1}^{L-1}\lambda_{2i}\sigma_i^z\sigma_{i+1}^z,
\eea
where $\sigma_i^{x,z}$ are the standard Pauli matrices acting on the $i$-th site of the chain.
Another representation for the fermionic case $N=2$, but for $p=2$, is given 
by the 3-spin interacting Fendley model \cite{fendley2}, where
\be \label{2.5}
H_M^{(2,2)} (\{\lambda_i\}) = - \sum_{i=1}^{M} \lambda_i\sigma_i^z\sigma_{i+1}^z
\sigma_{i+2}^x.
\ee

The general integrability condition for interacting multispin quantum chains, 
as presented in \cite{pozsgay1}, also explains the integrability 
of \rf{2.5} \cite{pozsgay2}.

A possible $Z(2)$ representation of \rf{2.2} and \rf{2.3} with arbitrary values of the 
integer parameter $p\geq 1$ is given by the $(p+1)$-interacting quantum 
chain
\be \label{2.6}
H_M^{(2,p)} (\{\lambda_i\}) = -\sum_{i=1}^M \lambda_i \sigma_i^z 
\sigma_{i+1}^z \cdots \sigma_{i+p-1}^z \sigma_{i+p}^x,
\ee
on a lattice with $M+p$ sites and open boundary conditions. 

For the $N>2$ cases the representations, due to \rf{2.2} and \rf{2.3}, are given  in 
terms of the $Z(N)$ generalized  Pauli matrices. They are 
$N\times N$ matrices satisfying
\be \label{2.7}
XZ = \omega ZX, \; X^N=Z^N=1,\; Z^\dagger = Z^{N-1},
\ee
and, as before, $\omega = \exp(i 2\pi/N)$. In the $Z$-basis they are given by the 
$Z(N)$-cyclic matrices
\be \label{2.8}
Z_{i,j} = \omega^{i-1}\delta_{i,j},\;  X_{i,j}= \delta_{j-i,1} + 
\delta_{i-j,N-1}; ~ i,j=1,\ldots,N. \non
\ee

The $Z(N)$ generalization of the Ising representation \rf{2.4} for $p=1$, gives 
us the free-parafermionic Baxter quantum chain \cite{baxter1}, with 
the Hamiltonian given by
\be \label{2.9}
H_{2L-1}^{(N,1)} = -\sum_{i=1}^L \lambda_{2i-1} X_i - \sum_{i=1}^{L-1} 
\lambda_{2i} Z_i Z_{i+1},
\ee
while the generalization of \rf{2.6}, for arbitrary $p$, leads to
\be \label{2.10}
H_M^{(N,p)} = -\sum_{i=1}^M \lambda_i Z_i Z_{i+1}\cdots Z_{i+p-1} X_{i+p}.
\ee

The non-vanishing eigenergies of all the Hamiltonians, \rf{2.4}-\rf{2.6},
 \rf{2.9} and \rf{2.10}, apart from degeneracies, are given by
\be \label{2.11}
E_{\scp{s_1,\ldots,s_{\mbar}}} = - \sum_{i=1}^{\mbar} \omega^{s_i} 
\varepsilon_i,
\ee
where $s_i$ takes {\it one and only one} of the possible values 
$s_i=0,1,\ldots,N-1$. The number $\mbar$ of quasi-energies $\varepsilon_i$ 
is given by
\be\label{2.12}
\overline{M} \equiv \mbox{int}\left(\frac{M+p}{p+1}\right)=\Big{\lfloor}{\frac{M+p}{p+1}}\Big{ \rfloor},
\ee
where hereafter we use $\inpar{a}$ to denote the integer part of $a$.

Each quasi-energy enters a single time in the composition of the eigenenergies 
\rf{2.11}. We can visualize the eigenenergies in \rf{2.11} as in Fig.~1, where 
we draw concentric circles of radius $\varepsilon_i$ in the complex plane.
 The intersections of the circles with the radial lines at angle 
$\frac{2\pi j}{N}$ 
($j=0,1,\ldots,N-1$) give us the possible contributions of the quasi-energies 
(open circles in Fig.~1) to the eigenenergies. Each circle gives {\it one and 
only one} contribution to \rf{2.11} (filled circles in Fig.~1).
\begin{figure} [htb]
\centering
\includegraphics[width=0.45\textwidth]{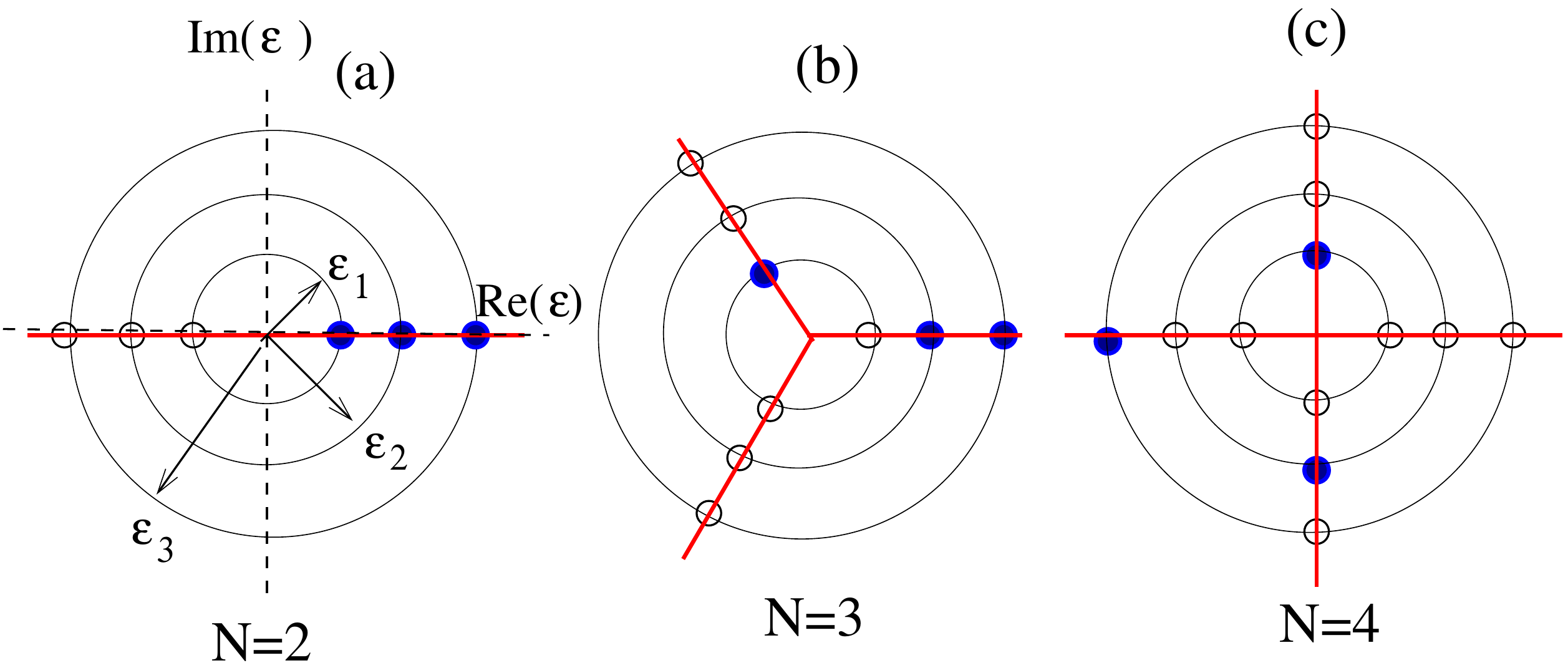}
\caption{  
Representation in the complex plane of the eigenenergies  in \rf{2.11} for 
the $Z(N)$ models with $N=2,3$ and 4. The circles have the radius $\varepsilon_i$, and the possible values (open circles) are the intercepts with the $Z(N)$-circles. Each circle contributes with  {\it one and only one} of the possible $N$ 
intercepts (black circles)}\label{fig1}
\end{figure}
This means that the quasi-particles $\varepsilon_i$ obey a $Z(N)$-circle 
exclusion that in the $N=2$ case is similar to the Pauli exclusion principle. 

The quasi-energies in \rf{2.11}
\be \label{2.22b}
\varepsilon_i=1/z_i^{1/N},
\ee
as shown in \cite{fendley2} for $p=N=2$, and for general values of $p$ and $N$ 
in \cite{AP1,AP2}, are obtained from the roots $z_i$ ($i=1,\ldots,\mbar$)
 of the polynomials 
\be \label{2.13}
P_M^{(p)}(z) = \sum_{\ell =0}^{\mbar} C_M(\ell)z^\ell, \quad 
P_M^{(p)}(z_i) =0, 
\ee
satisfying the recurrence relations 
\be \label{2.14a}
P_M^{(p)}(z) = P_{M-1}^{(p)}(z) -z\lambda_M^NP_{M-(p+1)}^{(p)}(z),
\ee
for $M\geq 1$, with the initial condition
\be \label{2.14b} 
P_M^{(p)}(z) =1, \mbox{ for } M\leq 0.
\ee
For real $\{\lambda_i\}$, we have checked that the roots of (\ref{2.13}) are real.

The ground state energy $E_0$ of the general $(N,p)$ models is real and it is 
obtained by taking $s_1=\cdots = s_{\mbar} =0$ in \rf{2.11},
\be \label{2.15}
E_0 = E_{\scp{0,0,\ldots,0}} = -\sum_{i=1}^{\mbar} \varepsilon_i.
\ee
While the ground state is real, the energy levels are in general complex.
We define the lower excited states as those with the lowest real part eigenvalue. In particular, the first excited states
are obtained
by taking the value $s_{\scp{1}} \neq 0$ and $s_k=0$ for $k=2,\dots, \bar M$ in (\ref{2.11}), that is,
$E_{\textrm{exc},j}=E_0 + (1-\omega^j) \varepsilon_1$ ($j=1,\ldots,N-1$). Accordingly,  the mass gap is defined
as,
\be \label{2.16}
\operatorname{Re}(\mbox{gap}_j) = (1- \cos(\frac{ 2 \pi j}{N})) \varepsilon_1; 
\quad j=1,\ldots,N-1.
\ee
In Fig.~1 the configuration for $N=2$ is the ground-state while the one for $N=3$ 
is one of the excited states that give the smallest gap.

In \cite{AP1,AP2} the $(N,p)$ models are shown to be critical at their 
isotropic point $\{\lambda_i=1\}$. The ground-state energy per site, at this 
critical point, is obtained analytically. It is given \cite{AP1} in terms of
the Lauricella hypergeometric series $F_D^{(p-1)}$ \cite{slater}. Also, the 
 gaps at these critical points have the finite-size leading dependence 
\be \label{2.17}
	\operatorname{Re}(\mbox{gap}) \sim \frac{1}{M^z}, \quad z= 
\frac{p+1}{M},
\ee
giving us the dynamical critical exponent $z$. The specific heat exponent 
$\alpha$, for $p=1$ is known analytically: $\alpha= 1-2/N$ \cite{AB1}. 
For $p>1$, the numerical solutions for the polynomial roots of 
\rf{2.14a}-\rf{2.14b}
for large lattice sizes, give us \cite{AP2}
\be \label{2.18}
\alpha = \max \{ 0,1-\frac{p+1}{N}\}.
\ee
 
\section{The XY quantum chains with multispin interactions}

We introduce in this section the general $U(1)$-symmetric 
XY models, which contain, in addition to 
nearest-neighbor interactions, $N$-multispin interactions 
($N=2,3,\ldots$). In their general form the Hamiltonians are given by
\bea \label{3.1}
&&H_M^{N,XY} (\{\mu_i\},\{\gamma_i\}) = \sum_{i=1}^{M+N-2} \mu_i \sigma_i^+ 
\sigma_{i+1}^-  \non \\
&&+ \sum_{i=1}^M \gamma_i \sigma_i^-\left( 
\prod_{j=i+1}^{i+N-2} \sigma_j^z \right) \sigma_{i+N-1}^+,
\eea
where $\sigma^{\pm} = (\sigma^x \pm i \sigma^y)/2$ and $\{\mu_i\},\{\gamma_i\}$  are real coupling constants. We can simplify the above Hamiltonian by performing the canonical 
transformation
\bea \label{3.2}
&&\sigma_1^{\pm} \rightarrow \sigma_1^{\pm};\,
\sigma_i^{\pm} \rightarrow \left( \prod_{j=1}^{i-1} \mu_j\right)^{\pm 1} 
\sigma_i^{\pm}, (i=2,\ldots,M+N-1), \non \\
&& \sigma_i^z \rightarrow \sigma_i^z \; (i=1,\ldots,M+N-1),
\eea
which gives
\bea \label{3.3}
H_M^{(N,XY)} (&&\{\lambda_i\}) =  \sum_{i=1}^{M+N-2} \sigma_i^+\sigma_{i+1}^- 
 \non \\
&&+\sum_{i=1}^M \lambda_i^N \sigma_i^- 
\left( \prod_{j=i+1}^{i+N-2} \sigma_j^z\right) \sigma_{i+N-1}^+,
\ee
where we have the effective parameters
\be \label{3.4}
\lambda_i^N = \gamma_i \left( \prod_{j=i}^{i+N-2} \mu_j\right). 
\ee
In Fig.~2 we draw the interactions in the Hamiltonian \rf{3.3} for some 
values of $N$.
\begin{figure} [htb]
\centering
\includegraphics[width=0.45\textwidth]{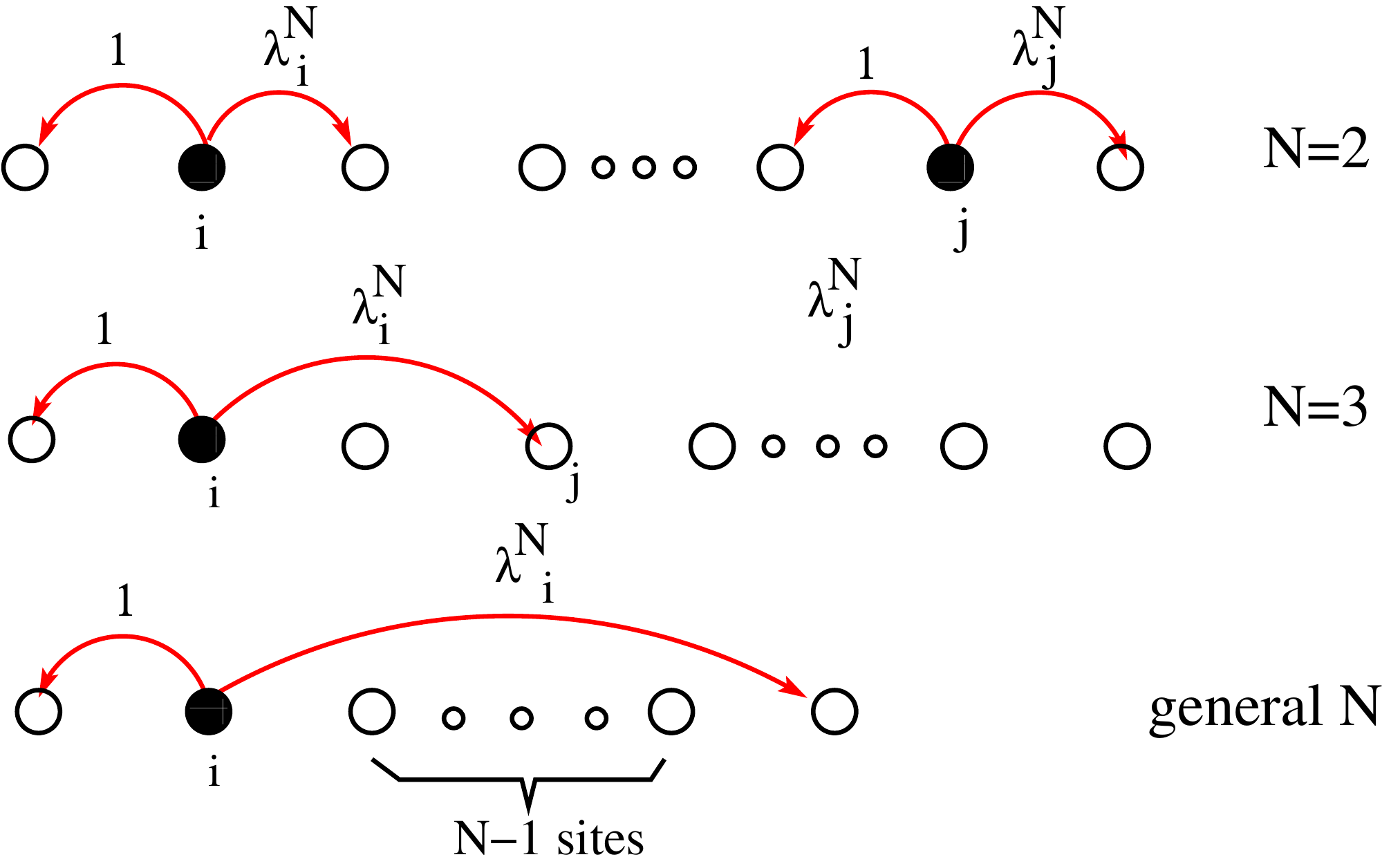}
\caption{  
The interactions in the Hamiltonian \rf{3.3}, for $N=2,3$ and for arbitrary 
$N$.}\label{fig2}
\end{figure}

We can verify that in the $N=2$ case, the Hamiltonian $H_M^{(2,XY)}$, after the 
canonical transformation
\bea \label{3.5}
&&\sigma_1^{\pm} \rightarrow \sigma_1^{\pm}, 
\sigma_i^{\pm} \rightarrow \left( \prod_{j=1}^{i-1} \lambda_j\right)^{\mp 1} 
\sigma_i^{\pm}, (i=2,\ldots,M+N-1), \non \\
&& \sigma_i^z \rightarrow \sigma_i^z \; (i=1,\ldots,M+1),
\eea
recovers the standard two-body non-uniform XY quantum chain 
(or XX quantum chain):
\be \label{3.6}
H_M^{(2,XY)} (\{\lambda_i\}) =  \sum_{i=1}^{M-1} 
\frac{\lambda_i}{2}(\sigma_i^x \sigma_{i+1}^x + \sigma_i^y\sigma_{i+1}^y),
\ee
having a real eigenspectrum due to its Hermiticity. For $N>2$, similarly to 
the $Z(N)$ free-parafermionic quantum chains of the preceding section, the Hamiltonians 
$H_M^{(N,XY)}$ are not Hermitian. However for all values of $N$, instead of 
being $Z(N)$ invariant the models have a larger $U(1)$ symmetry due to their 
commutations $[H_M^{(N,XY)},S^z]=0$, with the $z$-magnetization 
$S^z =\sum_i \sigma_i^z$. 

	In order to proceed let us introduce spinless fermionic operators 
through the Jordan-Wigner transformation \cite{lieb}
\be \label{3.7}
c_i = \sigma_i^-\prod_{j=1}^{i-1}\sigma_j^z, 
c_i^\dagger = \sigma_i^+\prod_{j=1}^{i-1}\sigma_j^z, 
\ee
for $i=1,\ldots, M+N-1$,
which satisfy the fermionic algebra
\be \label{3.8} 
 \{c_i,c_j^\dagger\} = \delta_{i,j},\quad \{c_i,c_j\}=0.
\ee
The Hamiltonian \rf{3.3}, in terms of these fermionic operators, has the 
bilinear form
\be \label{3.9}
H = -\sum_{i,j=1}^{M+N-1} c_i^\dagger \mathbb{A}_{i,j} c_j, 
\ee
where the connectivity matrix $A$ has the simple banded matrix form (tridiagonal for $N=2$):
\be \label{3.10}
\mathbb{A}_{i,j} = \delta_{j,i+1} + \lambda_j^N\delta_{j,i+1-N}.
\ee
Consider the  transformation $\{c_i,c_i^\dagger\} \to \{\eta_i,\eta_i^\dagger\}$:
\be \label{3.11}
\eta_k = \sum_i^{M+N-1}\mathbb{L}_{i,k} c_i, \quad 
\eta_k^\dagger = \sum_i^{M+N-1}\mathbb{R}_{i,k} c_i^\dagger,
\ee
where $\mathbb{L}_{i,j}$ and $\mathbb{R}_{i,j}$ are the components $i$ of the left and 
right eigenvector $j$ of the matrix $\mathbb{A}$, with eigenvalue $\Lambda_j$, i. e.,
\be \label{3.10a}
\sum_{k=1}^{M+N-1}\mathbb{A}_{i,k}\mathbb{R}_{k,j} = \Lambda_j \mathbb{R}_{i,j}, \quad 
\sum_{k=1}^{M+N-1}\mathbb{L}^T_{i,k}\mathbb{A}_{k,j} = \Lambda_j \mathbb{L}^T_{i,j}.\nonumber
\ee 
Since $\mathbb{A}$ is diagonalizable $\mathbb{L}^T \mathbb{R} =\id$, that also 
implies $\mathbb{R}\mathbb{L}^T=\id$. 
These relations and the fermionic relations \rf{3.8} imply that $\eta_k,\eta_k^\dagger$ 
are also fermionic operators:
\be\label{3.13}
\{\eta_k,\eta_{k'}^\dagger\} =\delta_{k,k'}, \quad \{\eta_k,\eta_{k'}\} =0.
\ee
Inverting \rf{3.11} we obtain
\be \label{3.13a}
c_i=\sum_{k=1}^{M+N-1} \mathbb{R}_{i,k} \eta_k, \quad c_i^\dagger = 
\sum_{k=1}^{M+N-1} \mathbb{L}_{i,k}\eta_k^{\dagger}.
\ee
Inserting \rf{3.13a} in \rf{3.9} we obtain
\be \label{3.11a}
H = -\sum_{k=1}^{M+N-1} \Lambda_k \eta_k^{\dagger} \eta_k.
\ee
This implies that all the $2^{M+N-1}$ eigenvalues of \rf{3.3} are obtained 
from the $M+N-1$ quasi-energies $\Lambda_k$ given by the eigenvalues of 
the matrix $A$, {\it i.e.},
\be \label{3.15}
E_{\scp{s_1,\ldots,s_{M+N-1}}} = -\sum_{k=1}^{M+N-1} \; s_k\Lambda_k; \; s_k =0,1 .
\ee
The quasi-energies $\Lambda_k$ are obtained by solving $\mbox{det}(A- 
\Lambda_k \id)=0$.
Apart from the zero modes ($\Lambda_k=0$), which will change the degeneracy 
 of the eigenspectrum, these quasi-energies are obtained from the zeros 
$\{z_k\}$ of the characteristic polynomial $P^{(N-1)}(z)$, given by
\be \label{3.16}
P_M^{(N-1)}(u) \equiv \mbox{Det}(\id -Au), 
P_M^{(N-1)}(u_k) =0, 
\eea
where $\Lambda_k = \frac{1}{u_k}$. 
 
These polynomials, as a consequence of the Laplace cofactor's rule for 
determinants, satisfy the recurrence relation 
\be \label{3.17}
P_M^{(N-1)}(z) = P_{M-1}^{(N-1)}(z) -z\lambda_M^N P_{M-N}^{(N-1)}(z),
\ee
where $z= (1/\Lambda)^N$, and 
\be \label{3.18}
P_M^{(N-1)} (z) = 1, \mbox{ for } M\leq 0.
\ee
From \rf{3.17} we can see that the order of the polynomial is 
$\inpar{(M+N-1)/N}$. Comparing \rf{3.17} and \rf{3.18} with 
\rf{2.14a} and \rf{2.14b}, we can see 
that the polynomials $P_M^{(N-1)}(z)$  are the same as those fixing the 
eigenspectra of the $Z(N)$ multispins chains with $N=p+1$ multispin 
interactions. The same roots $P_M^{(N-1)}(z_i)=0$ ($i=1,\ldots, 
\inpar{(M+N-1)/N})$ that give the quasi-energies $\varepsilon_i=1/z_i^{1/N}$ 
of the $Z(N)$ free-parafermionic multispin models also give us the 
quasi-energies of the fermionic XY chains with $N$-multispin interactions. 
Each root $z_i$ gives us $N$ fermionic quasi-energies 
\bea \label{3.19}
\Lambda_{j,i} = e^{i\frac{2 \pi}{N} j} \varepsilon_i,&& \quad i=1,\ldots,
\Big{\lfloor}{\frac{M+N-1}{N}}\Big{\rfloor}; \non \\
&&j=0,1,\ldots,N-1.
\eea
Since the total number of fermionic quasi-energies in \rf{3.15} is $M+N-1$, 
we should have
\be \label{3.20}
N_z = M+N-1 -N\Big{\lfloor}{\frac{N+M-1}{N}}\Big{\rfloor}
\ee
zero modes, producing a $2^{N_z}$-degeneracy in the whole eigenspectrum. 
Apart from this degeneracy the eigenlevels of the $N$-multispin interacting 
XY models \rf{3.3} are given by
\be \label{3.21}
E_{\{s_{i,j},r_{i,j}\}} =
-\sum_{i=1}^{\inpar{\frac{M+N-1}{N}}} 
\left( \sum_{j=0}^{N-1} r_{i,j}
\omega^{s_{i,j}}\right) \varepsilon_i, 
\ee
where for each $i=1,\ldots,\inpar{\frac{M+N-1}{N}}$, $s_{i,j}$  
takes one of the possible values $s_{i,j}=0,1,\ldots,N-1$ and 
$r_{i,j}=0,1$.

In the next section  we compare the eigenspectrum of the $Z(N)$ 
 free-parafermionic quantum chains and the $N$-multispin XY models.

\section{The correspondence of the XY model with $N$-multispin interactions 
and the $Z(N)$ free-parafermionic Baxter chain}

We show in this section that the eigenspectrum of the $N$-multispin 
interacting XY models, defined in the 
previous section, contains the eigenspectrum 
of the $Z(N)$ free-parafermionic Baxter quantum chains. This extends the 
known correspondence of the eigespectra of the standard Ising and XY quantum
chains. 

Let us consider initially the case $N=2$. By comparing \rf{3.17} and \rf{3.18} 
with 
\rf{2.14a} and \rf{2.14b}, we see that the polynomials $P_M^{(1)}(z)$, whose roots 
$\varepsilon_i=1/z_i^{1/2}$ fix the eigenspectra of the Ising and XY models 
are the same. This means from \rf{2.11} that, while a given quasi-energy 
contributes to the eigenlevels with $\varepsilon_i$ or $-\varepsilon_i$, 
in the Ising chain, the contribution in the XY model can be of four 
possible values: $0,-\varepsilon_1,+\varepsilon_1,
\varepsilon_1-\varepsilon_1$ $\equiv 
 -\varepsilon_1,0,0,+\varepsilon_1$. In Fig.~3 we show pictorially the 
contribution of a given quasi-energy $\varepsilon_i$ to the eigenenergies 
of both models.
\begin{figure} [htb]
\centering
\includegraphics[width=0.45\textwidth]{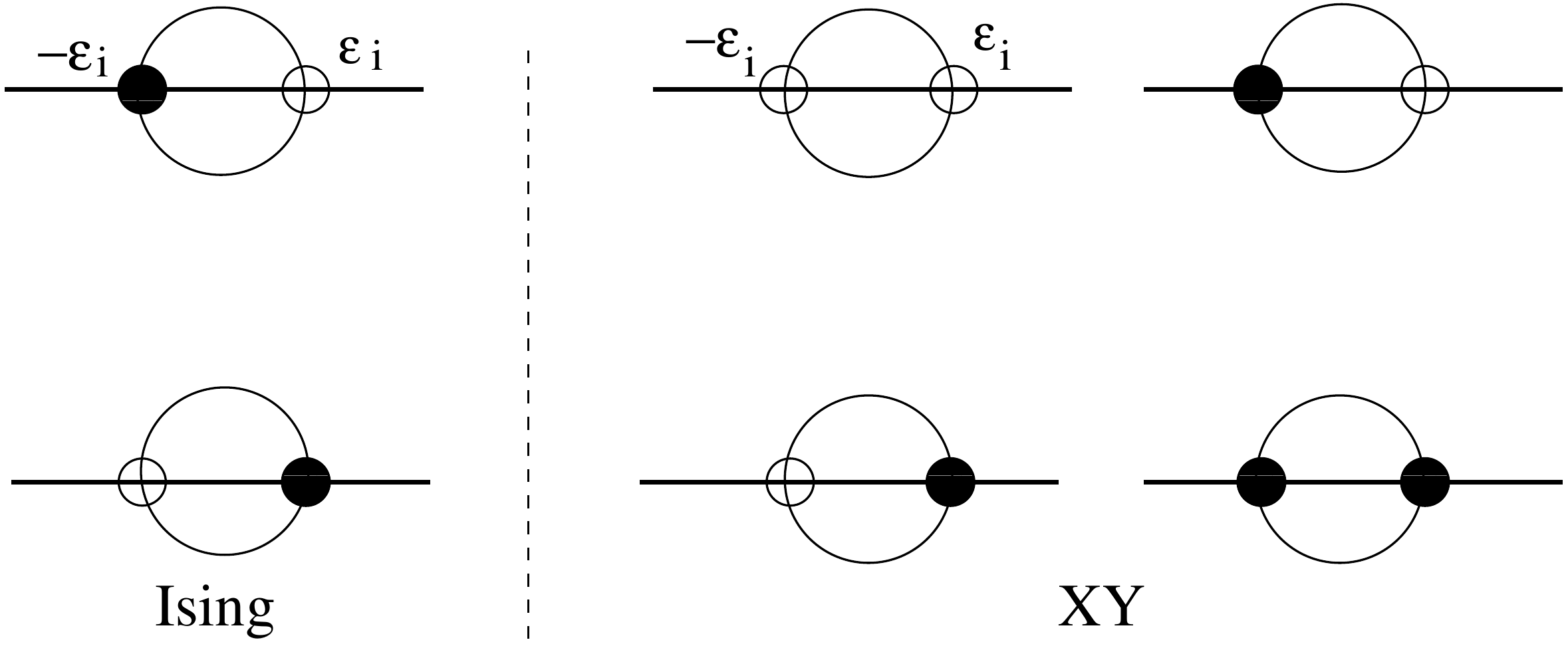}
\caption{  
The two and four possible contributions of the quasi-energy $\varepsilon_i$ for the Ising chains and 
for the XY model, respectively.} \label{fig3}
\end{figure}
We can see from the figure that while in the Ising case we have a $Z(2)$ circle 
exclusion for the quasi-energy, in the XY model there is no such exclusion. 
An independent check of this correspondence comes from the algebraic 
properties of $\sigma_i^x\sigma_{i+1}^x$ and $\sigma_i^y\sigma_{i+1}^y$ 
operators in \rf{3.6}. The XY model \rf{3.6} and two decoupled Ising 
models with coupling $\{\lambda_i/2\}$ (see \cite{abb}) are given in 
terms of density-energy operators satisfying the same algebraic rules. This 
implies that 
the contributions of the eigenenergies are $(-\varepsilon_i/2,
\varepsilon_i/2)\oplus (-\varepsilon_i/2,\varepsilon_i/2)$ i. e.,
($-\varepsilon_i,0,0,\varepsilon_i$).

Let us now consider the general cases $N\geq 2$. We split the $M$ 
coupling constants ($\lambda_1,\ldots,\lambda_M$) in cells of size $N$:

\be \label{4.1}
M = N \Big{\lfloor}{\frac{M}{N}}\Big{ \rfloor} + \ell_M
\ee
and in each cell only the two first coupling constants are nonzero, \textit{i.e.},
\be \label{4.2}
\lambda_i=\lambda_{(k-1)N+j}=0, \mbox{ for } j=3,\ldots,N.\ee
We redefine the nonzero coupling constants 
\be \label{4.3}
\tilde{\lambda}_{\ell} = \tilde{\lambda}_{2(k-1)+j} \equiv \lambda_{(k-1)N+j} 
\mbox{ for } j=1,2,
\ee
so that $\ell = 1,2,\ldots,\widetilde{M}$, and the number of nonzero 
$N$-multispin couplings in the chain is
\be \label{4.4}
\widetilde{M} = 2\Big{\lfloor}{\frac{M}{N}}\Big{\rfloor} + \mbox{min}(\ell_M,2),
\ee
while $\ell_M$ is given in \rf{4.1}. The set of coupling constants, from 
\rf{4.2}, are $\{\tilde{\lambda}_1^N,\tilde{\lambda}_2^N,0,\cdots,0; 
\tilde{\lambda}_3^N,\tilde{\lambda}_4^N,0,\cdots,0;\cdots \}$. 
The $N$-multispin XY model \rf{3.3} is now given by
\bea \label{4.5}
H_{N,2}^{(N,XY)} &&(\{\tilde{\lambda}_i\}) =  \sum_{i=1}^{M+N-2} \sigma_i^+\sigma_{i+1}^- 
 \non \\
&&+\sum_{\ell=1}^{\widetilde{M}} \tilde{\lambda}_{\ell}^N \sigma_{\tilde{\ell}}^- 
\left( \prod_{j=\tilde{\ell}+1}^{\tilde{\ell}+N-2} \sigma_j^z\right) 
\sigma_{\tilde{\ell}+N-1}^+,
\ee
where 
\be \label{4.6}
\tilde{\ell} = \ell + (N-2)
\Big{\lfloor}{\frac{\ell-1}{2}}\Big{ \rfloor}.
\ee
For example for $N=3$ and $M=4$ we have 
\bea \label{4.7}
&&H_{3,2}^{(3,XY)} = \sum_{i=1}^5 
\sigma_i^+\sigma_{i+1}^- 
+\tilde{\lambda_1}^3 \sigma_1^-\sigma_2^z\sigma_3^+ 
+\tilde{\lambda_2}^3 \sigma_2^-\sigma_3^z\sigma_4^+ \non \\ 
&&_+\tilde{\lambda_3}^3 \sigma_4^-\sigma_5^z\sigma_6^+.
\eea

To proceed, let us insert \rf{4.2} in the recurrence relation 
\rf{3.17}, and we obtain for $j=3,\ldots,N$
\be \label{4.8}
P_i^{(N-1)} = P_{(k-1)N+j}^{(N-1)} = P_{(k-2)N+2}^{(N-1)},
\ee
and for $j=1,2$ we define similarly as in \rf{4.3}
\be \label{4.9}
\widetilde{P}_{\ell}^{(1)} = P_i^{(N-1)} = P_{(k+1)N+j}^{(N-1)}, 
\ee
with $\ell =(k-1)2+j$. It follows from the above relations that 
\be \label{4.10} 
P_{i-1}^{(N-1)} \equiv \widetilde{P}_{\ell-1}^{(1)},
\ee
\be \label{4.11}
P_{i-1}^{(N-1)} = P_{(k-2)N +j}^{(N-1)} = \widetilde{P}_{(k-2)2+j}^{(1)} 
\equiv 
\widetilde{P}_{\ell-2}^{(1)}
\ee
Inserting \rf{4.10} and \rf{4.11} in the recurrence relations \rf{3.17} and \rf{3.18},
 we obtain 
\be \label{4.12}
P_{\ell}^{(1)}(z) = \widetilde{P}_{(\ell-1)}^{(1)}(z) 
-z\tilde{\lambda}_{\ell}^N
\widetilde{P}_{\ell-2}^{(1)}(z), 
\ee
for $\ell =1,\ldots,\widetilde{M}$, with the initial condition 
\be \label{4.13}
\widetilde{P}_{\ell}^{(1)}(z) =1, \mbox{ for } \ell \leq 0.
\ee
The above recurrence relations are precisely the ones for the polynomial 
$P_M^{(p)}(z)$ given in \rf{2.14a} and \rf{2.14b} for the $Z(N)$ multispin 
chains with $p=1$ and the identification $\widetilde{M} \leftrightarrow M$ and 
$\tilde{\lambda}_{\ell} \leftrightarrow \lambda_{\ell}$. The roots $z_i$ of 
these polynomials give us the quasi-energies $\varepsilon_i = 1/z_i^{1/N}$ 
of the models. This implies that the $N$-multispin XY model with Hamiltonian 
$H_{N,2}^{(N,XY)}$ given in \rf{4.6} and \rf{4.7} 
 has the same quasi-energies as the 
$Z(N)$ multispin free-parafermionic Hamiltonian $H_{\widetilde{M}}^{(N,1)}$ 
given in \rf{2.10}.
In the particular case where $\widetilde{M}=2L-1$ is an odd 
number, the XY model \rf{4.6} is related to the $Z(N)$ free-parafermionic 
Baxter chain, with the  Hamiltonian given in \rf{2.9}. The quasi-energies 
$\varepsilon_i$ ($i=1,\ldots,L$), from \rf{3.21} give the eigenspectra of 
$H_{N,2}^{(N,XY)}$ 
\be \label{4.14}
E_{\{s_{i,j},r_{i,j}\}} = -\sum_{i=1}^L\left( \sum_{j=0}^{N-1} r_{i,j}
\omega^{s_{i,j}}\right) \varepsilon_i,
\ee
where $s_{i,j}=0,1\ldots,N-1$ and $r_{i,j}=0,1$. From   \rf{2.11} 
the same quasi-energies also give  the eigenspectra of $H_{2L-1}^{(N,1)}$
\be \label{4.15}
E_{s_1,\ldots,s_L} = - \sum_{i=1}^L \omega^{s_i} \varepsilon_i,
\ee
where $s_i=0,1,\ldots,N-1$ and $\omega=\exp(i2\pi/N)$.

In \rf{4.15} a given quasi-energy $\varepsilon_i$ contributes to an eigenlevel 
with one of the $N$ possible values $\omega^{s_i}\varepsilon_i$ ($s_i=0,1,\ldots,N-1$), 
 while in \rf{4.14} the same root contributes with $2^N$ distinct values. 
In Fig.~4 we draw pictorially the contributions of a given root 
$\varepsilon_i$ to both models with $N=3$.
\begin{figure} [htb]
\centering
\includegraphics[width=0.45\textwidth]{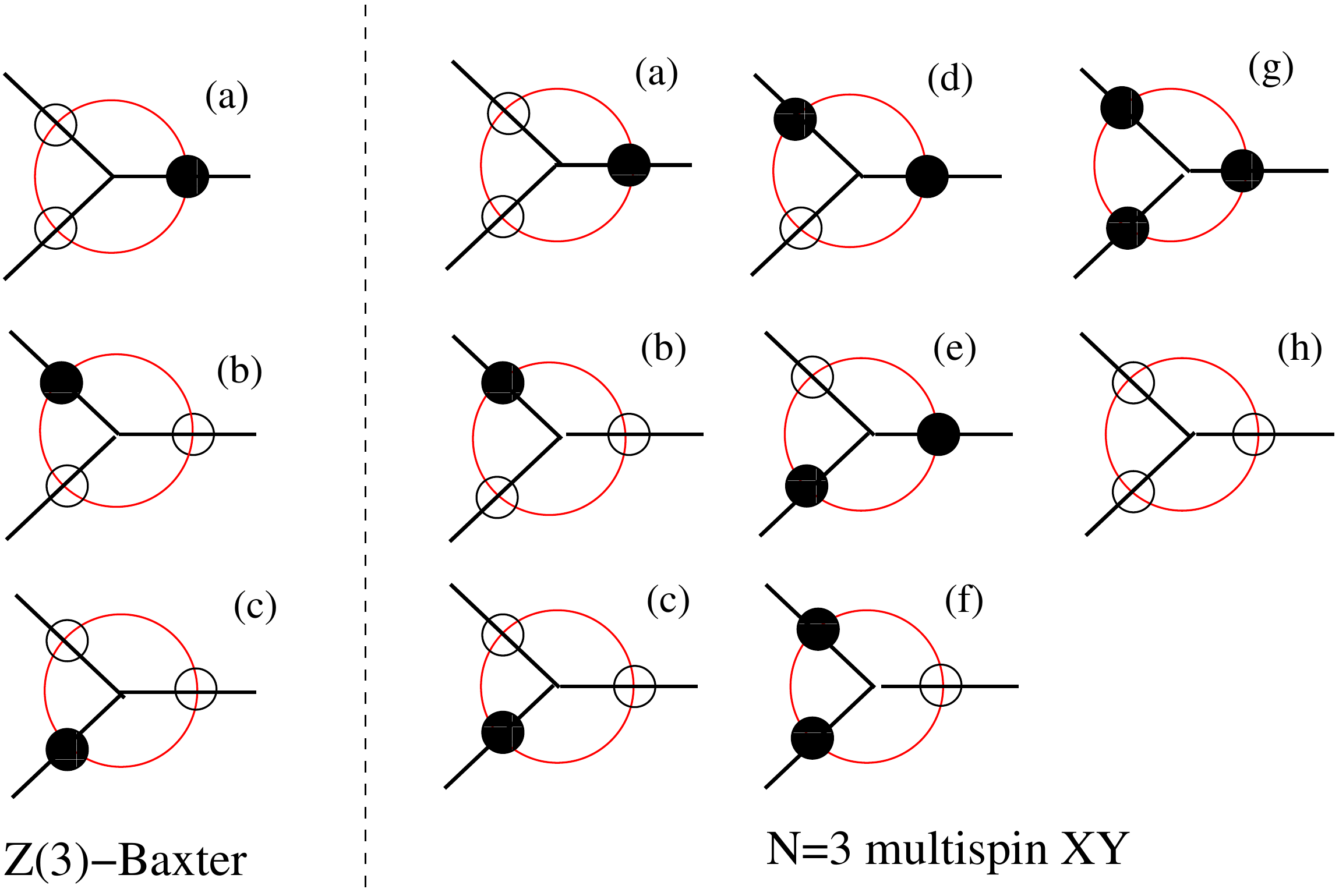}
\caption{  
The possible contributions of the quasi-energy $\varepsilon_i$ for the $Z(3)$ 
free-parafermionic Baxter chain and for the equivalent XY model with 
$N$-multispin interactions. In the $Z(3)$ model the quasi-energy enters 
only one time (circle repulsion), differently from the related XY 
model where we can have all the possible values (no circle repulsion).} 
\label{fig4}
\end{figure}
In the free-parafermionic $Z(N)$ chain (see Fig.~4) we have a $Z(N)$ circle 
repulsion for the quasi-energies, that is absent in the  $N$-multispin XY 
quantum chain. These results generalize the correspondence of the 
Ising-XY models \cite{abb} (see Fig.~3) to the $Z(N)$ free-parafermionic models.

All the eigenenergies of the $Z(N)$ free-parafermion model also appear in 
the $U(1)$ sector of the XY model with $L$ particles. In particular, the 
ground-state energy and the mass gap are the same. In Fig.~5 
 we show for the model with symmetry $Z(4)$  
the quasi-energies forming the ground state and one of the lowest eigenenergy 
states of both models. 
In Appendix B, we show a numerical comparison of the quasi-energies 
of the $N=4$ multispin XY and the $Z(4)$ Baxter chain.
\begin{figure} [htb]
\centering
\includegraphics[width=0.45\textwidth]{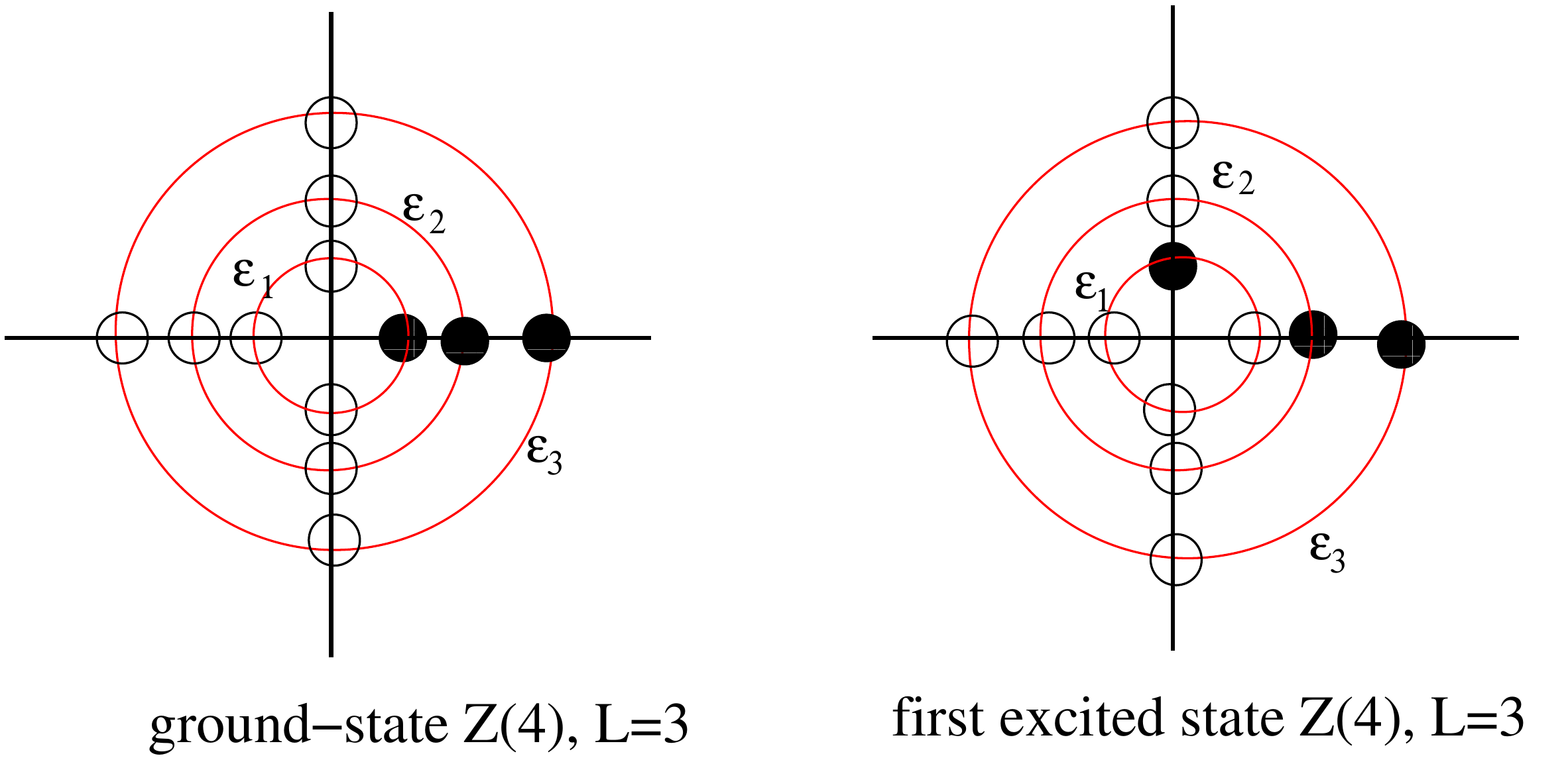}
\caption{  
Contribution of the quasi-energies $\varepsilon_i$ to the ground state  
and one of the first-exited states for the $Z(4)$ free-fermionic model and
for the $N=4$-multispin interactions in a lattice size $L=3$.} \label{fig5}
\end{figure}

From the exact results of the $Z(N)$ free-parafermionic chains, at the 
critical isotropic point $\{\tilde{\lambda}_i=1\}$ \cite{AP1,AP2} 
the dynamical and specific heat exponents for the XY model \rf{4.5} are 
given by \rf{2.17} and \rf{2.18}, respectively. 

We can also obtain the XY models with multispin interactions that share the 
quasi-energies with the $Z(N)$ free-parafemionic quantum chains with $(p+1)$ 
interactions introduced in \cite{AP1,AP2}. The construction follows from a 
generalization of the procedure we just presented. In Appendix A, 
we present 
these models. The construction of the equivalent XY models we present, only 
works for the models where $N\geq p+1$. Differently from the case $p=1$, where we can derive the $N$-XY model for arbitrary $Z(N)$ symmetry, in the cases 
where $p>1$ we only found the correspondence for $N\geq p+1$. This means, 
for example,  that 
for the multispin model with $p=2$ we only found the related $N$-XY models 
with $N=3,4,\ldots$. This excludes, for example, the $Z(2)$ fermionic 
3-spin interacting Fendley model \cite{fendley2}.
In Appendix B, we show a numerical comparison of the quasi-energies 
of the $N=4$ multispin XY and the $Z(4)$ free-parafermionic model 
with 3-multispin interactions $(p=2)$.

\section{Conclusions}

The relationship between the standard Ising model and XY quantum chains is well 
known. The eigenergies of the XY quantum chain can be obtained from the ones 
of two decoupled Ising Hamiltonians. Both models have a free-fermionic 
eigenspectra. In this paper we show that the eigenspectra of several quantum 
chains with a free-parafermionic quasi-particle eigenspectra can also be 
obtained from generalized XY quantum chains. These XY models for $N=2$ recover the standard XY quantum chain. 
For $N>2$ they are non-Hermitian, like the known free-parafermionic $Z(N)$ 
models. These generalized XY models contain, in addition to two-body interactions, also 
$N$-multispin interactions.  

The $Z(N)$ symmetry in the free-parafermionic chains is enhanced to an 
$U(1)$ symmetry in the generalized XY models. The spectrum of the generalized XY model is given in terms of linear
combinations of all the
quasi-energies of the free parafermionic Z(N) models; see Eq.(\ref{4.14}). 
In particular, the special $U(1)$ sector with $\mbar$ fermions contains 
all the
$N^{\mbar}$ eigenlevels of the corresponding free parafermionic model.
The ground-state and the low-lying excited states have the same energy in both models
for $N \leq 4$, implying that at their critical point they share the same values
for the critical exponents. For $N>4$, differently from the $Z(N)$
free-parafermionic models, where the ground-state is always real, in the related
XY models the eigenenergy with the lowest real part is complex.

Since the exact values for some of the exponents are 
known for the $Z(N)$ models, they are also exact for these generalized 
XY models. 

The equivalence we found in this paper is also important to better 
understand the exact integrability of the $Z(N)$ free-parafermionic models. 
The exact solution of the $N$-multispin XY models comes from a standard 
Jordan-Wigner transformation, quite differently from the exact solution of the 
$Z(N)$ 
free-parafermionic quantum chain. 

The closed solution of all the $Z(N)$ free-parafermionic models was obtained 
only in the case of free boundary conditions. In the periodic case, 
although the model is still exactly integrable, the solution is more complicated 
and the eigenspectrum is not a free-parafermionic one. It probably has an almost 
free particle eigenspectrum, but with some selection rules. On the other hand 
the related XY models remains with a free-fermionic eigenspectrum even 
for the periodic case. This implies that the equivalence shown in this paper 
is restricted to the free-boundary case. 

The free-parafermionic $Z(N)$ Baxter models ($p=1$), for $N\geq3$ show, in 
the thermodynamic limit,  
an anomalous behavior for the ground-state energy per site \cite{AB2}. 
Usually this quantity does not depend on the imposed boundary condition. 
However. for the  $Z(N)$ models this quantity is distinct for the 
periodic and free-boundary conditions. It will be interesting to verify if 
this anomaly is also present in the related $N$-multispin XY model. 

Finally we should mention that we did not find the equivalent multispin XY  
quantum chain related to the $Z(N)$ parafermionic models with $(p+1)$ 
interactions when $N<(p+1)$. The existence of such XY models is certainly 
an interesting question to be answered in the future. 

\begin{acknowledgments}
We gratefully acknowledge
 discussions with Jos\'e A. Hoyos.
This work was supported  in part by the Brazilian agencies FAPESP and CNPq. RAP was supported by CNPq (Grant No. 150829/2020-5).
\end{acknowledgments}

\appendix

\section{The correspondence of the generalized XY model and the $Z(N)$ free-parafermionic models with $(p+1)$-multispin interactions}

In this appendix we generalize the results of Sec.~4 , to obtain the $N$-XY 
models that contain the eigenspectra of the $Z(N)$ free-parafermionic 
quantum chains with $(p+1)$-multispin interactions \cite{AP1,AP2}. The 
generalization we present only works for $N\geq p+1$.

As in \rf{4.1}, we split the $M$ coupling constants of \rf{3.3} 
$\{\lambda_1,\ldots,\lambda_M\}$ in cells containing $N$ lattice sites. In 
each cell we cancel the coupling constants 
\be \label{5.1}
\lambda_i=\lambda_{(k-1)N+j} =0, \mbox{ for } j=p+2,\ldots,N,
\ee
and we define
\be \label{5.2}
\tilde{\lambda}_{\ell} = \tilde{\lambda}_{(p+1)(k-1)+j} \equiv 
\lambda_{(k-1)N +j},  \mbox{ for } j=1,\ldots,p+1, \non
\ee
so that $\ell = 1,2,\ldots,\widetilde{M}$, and the number of non-zero 
couplings is
\be \label{5.3}
\widetilde{M} = (p+1){\Big{\lfloor}}{\frac{M}{N}}{\Big{\rfloor}} + \mbox{min}(\ell_M,p+1),
\ee
with $\ell_M$ given in \rf{4.1}. The $N$-multispin XY model \rf{3.3} is now
given by
\bea \label{5.4}
H_{N,p+1}^{(N,XY)} &&(\{\tilde{\lambda}_i\}) =  \sum_{i=1}^{M+N-2} \sigma_i^+\sigma_{i+1}^- 
 \non \\
&&+\sum_{\ell=1}^{\widetilde{M}} \tilde{\lambda}_{\ell}^N \sigma_{\tilde{\ell}}^- 
\left( \prod_{j=\tilde{\ell}+1}^{\tilde{\ell}+N-2} \sigma_j^z\right) 
\sigma_{\tilde{\ell}+N-1}^+,
\ee
where 
\be \label{5.5}
\tilde{\ell} = \ell + (N-p-1)
\Big{\lfloor}{\frac{\ell-1}{p+1}}\Big{ \rfloor}.
\ee
As an example, for $N=4$, $M=6$ and $p=2$, we have
\bea \label{5.6}
&&H_{4,3}^{(4,XY)} = \sum_{i=1}^{8} 
\sigma_i^+\sigma_{i+1}^- 
+\tilde{\lambda_1}^4 \sigma_1^-\sigma_2^z\sigma_3^z\sigma_4^+ \non \\
&&+\tilde{\lambda_2}^4 \sigma_2^-\sigma_3^z\sigma_4^z\sigma_5^+
+\tilde{\lambda_3}^4 \sigma_3^-\sigma_4^z\sigma_5^z\sigma_6^+
+\tilde{\lambda_4}^4 \sigma_5^-\sigma_6^z\sigma_7^z\sigma_8^+ \non \\
&&+\tilde{\lambda_5}^4 \sigma_6^-\sigma_7^z\sigma_8^z\sigma_9^+. 
\eea

Inserting \rf{5.1} in the recurrence relations \rf{3.17} we obtain for
$j=p+2,\ldots,N$
\be \label{5.7}
P_i^{(N-1)} = P_{(k-1)N+j}^{(N-1)} = P_{(k-2)N+p+1}^{(N-1)},
\ee
and for $j=1,\ldots,p+1$ we define 
\be \label{5.8}
\widetilde{P}_{\ell}^{(p)} = P_i^{(N-1)} = P_{(k+1)N+j}^{(N-1)}, 
\ee
with $\ell =(k-1)(p+1)+j$. 
From  the above relations we obtain the generalization of \rf{4.10}-\rf{4.11}:
\be \label{5.9} 
P_{i-1}^{(N-1)} \equiv \widetilde{P}_{\ell-1}^{(p)},
\ee
\be \label{5.10}
P_{i-1}^{(N-1)} = P_{(k-2)N +j}^{(N-1)}) = \widetilde{P}_{(k-2)(p+1)+j}^{(p)} 
\equiv 
\widetilde{P}_{\ell-(p+1)}^{(p)}
\ee
Inserting the last relations in the recurrence \rf{3.17}, we finally obtain
\be \label{5.11}
\tilde{P}_{\ell}^{(p)}(z) = \widetilde{P}_{\ell-1}^{(p)}(z) -z\tilde{\lambda}_{\ell}^N
\widetilde{P}_{\ell-(p+1)}^{(p)}(z), 
\ee
for $\ell =1,\ldots,\widetilde{M}$, with the initial condition 
\be \label{5.12}
\widetilde{P}_{\ell}^{(p)}(z) =1, \mbox{ for } \ell \leq 0.
\ee
These relations are the same as the ones of the $Z(N)$ $(p+1)$-multispin 
free-parafermionic models given in \rf{2.14a}-\rf{2.14b}. This means that 
the XY  model \rf{5.4}-\rf{5.5} and \rf{2.10} are given by a free-particle 
eigenspectrum with the same quasi-energies.  

\section{Comparison of quasi-energies of the $Z(N)$ parafermionic chains and 
the $N$-multispin chains for small lattice sizes}
\begin{table*}[t]
\centering
\begin{tabular}{|cc|c||cc|c|}
\hline 
XY ($N=4,p=1$)& & $Z(4)$ Baxter & XY ($N=4,p=2$)&& 3-spin $Z(4)$ \\ \hline \hline
3.139634 +0I =$\varepsilon_1$  & 0.955525 +0I = $\varepsilon_2$ & 
3.139634=$\varepsilon_1$ &
4.335391+0I =$\varepsilon_1$  &
0.922639+0I =$\varepsilon_2$  & 4.33539 =$\varepsilon_1$ \\ \hline
0+3.139634I =$\varepsilon_1\omega$  & 0+0.955525I  = $\varepsilon_2\omega$ & 
0.9555253=$\varepsilon_2$ &
0+4.335391I =$\varepsilon_1\omega$  &
0+0.922639I =$\varepsilon_2\omega$  & 0.922639 =$\varepsilon_2$ \\ \hline
-3.139634+0I =$\varepsilon_1\omega^2$  & -0.955525 +0I  = $\varepsilon_2\omega^2$ & - &
-4.335391 +0I =$\varepsilon_1\omega^2$  &
-0.922639I +0I=$\varepsilon_2\omega^2$  & -  \\ \hline
0 -3.139634I =$\varepsilon_1\omega^3$  & 0 -0.955525I  = $\varepsilon_2\omega^3$ & - &
0 -4.335391I =$\varepsilon_1\omega^3$  &
0 -0.922639I=$\varepsilon_2\omega^3$  & -  \\ \hline
\end{tabular}
\caption{Quasi-energies for the related free-fermionic $N$-multispin XY and 
the free-parafermionic $Z(N)$ quantum chains, for $M=5$, $N=4$ and 
$\omega=\exp(i2\pi/4)$. 
The quasi-energies for the 
XY models are the ones of the Hamiltonian $H_{N,p+1}^{N,XY}$ given in 
\rf{5.4},and for the  free-parafermionic $Z(N)$ cases they are the ones
of  the  Baxter \rf{2.9} and 3-spin interaction chains \rf{2.10}.}
\label{table1}
\end{table*}

In this appendix, we give two simple examples of the quasi-eigenenergies of 
the $Z(N)$ free-parafermionic with  $(p+1)$-multispin interaction and the $N$-multispin 
free-fermionic XY quantum chains. In both examples we consider 
  $M=5$, $N=4$.

In the first example  the non-zero coupling constants  are
\rf{4.3}:
\be \label{b1}
\tilde{\lambda}_1=\lambda_1 =1,  
\tilde{\lambda}_2=\lambda_2 =2,  
\tilde{\lambda}_3=\lambda_5 =3.  
\ee
In this case the XY model is (see \rf{4.5}) $H_{4,2}^{(4,XY)}(\tilde{\lambda}_1,
\tilde{\lambda}_2,\tilde{\lambda}_3)$. Its quasi-energies are related to the 
ones in the $Z(4)$-Baxter chain \rf{2.9}. From \rf{4.4} the number of 
density operators in the Hamiltonian is $\widetilde{M}=3$, and it corresponds 
to the $Z(4)$ Baxter chain \rf{2.9} with $L= (\widetilde{M}+1)/2=2 $ sites, 
with the coupling constants
$\tilde{\lambda}_1,\tilde{\lambda}_2$ and $\tilde{\lambda}_3$.  The polynomial 
\rf{4.12}-\rf{4.13} is given by
\be \label{b2}
\widetilde{P}_3^{(1)}(z) = 1 -
(\tilde{\lambda}_1^N +
\tilde{\lambda}_2^N +
\tilde{\lambda}_3^N )z
+ \tilde{\lambda}_1^N\tilde{\lambda}_3^N z^2.
\ee
The roots $z_{\pm}$ of this polynomial give the quasi-energies
\be \label{b3}
\varepsilon_1 = \frac{1}{z_{+}^{1/4}}=3.139634, \;
\varepsilon_2 = \frac{1}{z_{-}^{1/4}}=0.955525 .
\ee
We show in the first three columns of Table I  the numerical values of the quasi-energies appearing in
the XY and $Z(4)$ related quantum chains.

The second example is for the multispin XY model related to the $Z(4)$
free-parafermionic quantum chain with 3-multispin interactions ($p=2$).
The non-zero coupling constants in
 \rf{3.3} are now:
\be \label{b4}
\tilde{\lambda}_1=\lambda_1 =1,
\tilde{\lambda}_2=\lambda_2 =2,
\tilde{\lambda}_3=\lambda_3 =3,
\tilde{\lambda}_4=\lambda_5 =4. \non
\ee
The XY model is now (see appendix A),
$H_{4,3}^{(4,XY)} (\tilde{\lambda}_1,\tilde{\lambda}_2,\tilde{\lambda}_3,\tilde{\lambda}_4)$,
with $\widetilde{M}=4$. The  polynomial \rf{5.11}-\rf{5.12} is
of degree $\inpar{(\widetilde{M} +p)/(p+1)}=2$
\be \label{b5}
\widetilde{P}_4^{(2)}(z) = 1 -
(\tilde{\lambda}_1^N +
\tilde{\lambda}_2^N +
\tilde{\lambda}_3^N +
\tilde{\lambda}_4^N )z
+ \tilde{\lambda}_1^N\tilde{\lambda}_4^N z^2,\non
\ee
with roots $z_{\pm}$, giving us the quasi-energies
\be \label{b6}
\varepsilon_1 = \frac{1}{z_{+}^{1/4}}=4.335391, \;
\varepsilon_2 = \frac{1}{z_{-}^{1/4}}=0.922639 .
\ee
In the last three columns of Table I we show the numerical values of the quasi-energies in the
XY and $Z(4)$ quantum chains with 3-spin interactions.


\end{document}